\documentclass[10pt, conference, compsocconf]{IEEEtran}
\ifCLASSINFOpdf
  \usepackage[pdftex]{graphicx}
\else
\fi
\usepackage{url}
\makeatletter
\def\url@leostyle{%
  \@ifundefined{selectfont}{\def\UrlFont{\sf}}{\def\UrlFont{\small\ttfamily}}}
\makeatother
\newcommand{\bi}{\begin{itemize}} 
\newcommand{\ei}{\end{itemize}} 
\newcommand{\ii}{\item} 

\begin{document}
%
\title{The use of microblogging for field-based scientific research}

\author{\IEEEauthorblockN{Alberto Pepe}
\IEEEauthorblockA{Center for Astrophysics \&\\
Institute for Quantitative Social Science\\
Harvard University\\
Cambridge, MA, USA\\
\texttt{apepe@cfa.harvard.edu}}
\and
\IEEEauthorblockN{Matthew S. Mayernik}
\IEEEauthorblockA{NCAR Library\\
National Center for Atmospheric Research\\
Boulder, CO, USA\\
\texttt{mayernik@ucar.edu}}
}


%


\maketitle

\begin{abstract}
Documenting the context in which data are collected is an integral
part of the scientific research lifecycle. In field-based research,
contextual information provides a detailed description of scientific
practices and thus enables data interpretation and reuse. For field
data, losing contextual information often means losing the data
altogether. Yet, documenting the context of distributed,
collaborative, field-based research can be a significant challenge due
to the unpredictable nature of real-world settings and to the high
degree of variability in data collection methods and scientific
practices of different researchers. In this article, we propose the
use of microblogging as a mechanism to support collection, ingestion,
and publication of contextual information about the variegated digital
artifacts that are produced in field research. We perform interviews
with scholars involved in field-based environmental and urban sensing
research, to determine the extent of adoption of Twitter and similar
microblogging platforms and their potential use for field-specific
research applications. Based on the results of these interviews as
well as participant observation of field activities, we present the
design, development, and pilot evaluation of a microblogging
application integrated with an existing data collection platform on a
handheld device. We investigate whether microblogging accommodates the
variable and unpredictable nature of highly mobile research and
whether  it represents a suitable mechanism to document the context of
field research data early in the scientific information lifecycle.
\end{abstract}


%
\IEEEpeerreviewmaketitle

\section{The context of field research data}
Digital technologies, such as sensor networks in the environmental sciences, social networking tools in the social sciences, and the digitization of cultural artifacts in the humanities, allow researchers to produce digital research products with volumes and complexities much greater than were possible in the past. This deluge of digital artifacts offers tremendous opportunity for new kinds of research \cite{borgman_book}. The integration of many diverse data streams from numerous small-scale projects can enable synthesis and longitudinal studies of large-scale environmental phenomena such as climate change and species shifts. Similarly, the digital integration of historical texts, photos, and maps from libraries and archives all over the world can enable comparative humanities projects that would otherwise be difficult or impossible to conduct. But while the possibilities of these kinds of digital resources are exciting, they also present tremendous challenges. Scientific artifacts that originate in digital form exist in a wide array of genres, including manuscripts, publications, data, laboratory and field notes, instrument calibrations, preprints, grant proposals, talks, slides, patent applications, theses, and dissertations, among others. This ensemble of artifacts and the relationships among them can be conceptualized as a lifecycle \cite{pepe:ore}. Artifacts that originate for one purpose at a stage of the lifecycle may be sought later for other purposes and at other stages, or they may be in active use for some time and then lay dormant or be discarded. 

Increasingly, scientific artifacts such as data and articles are being produced in collaborative research settings. As research becomes more distributed and dependent on digital means for collection and dissemination of research products, collections of digital resources, in order to be useful outside local settings, must be supplemented by information about the resources themselves. Understanding the context in which data and documents were produced, and the ways that they were modified and analyzed, is essential if they are to be used, shared, and re-used. This ``information about data'' is commonly referred to as ``metadata,'' though the term metadata often connotes formalized description and representation schemes.  In distributed scientific collaborations, however, creating and collecting contextual information about digital artifacts is a significant challenge due to the high degree of variability in data collection methods, data types, and data uses. Contextual information about scientific artifacts, or ``contextual data", are often shared informally as well as formally in scientific collaborations \cite{edwards}.

In this paper, we focus on the challenge of capturing and collecting contextual information in dedicated scientific data repositories. This challenge is compounded by the various and amorphous conceptions of ``context'' itself. Paul Dourish provides some light on the topic in an analysis of differing conceptions of ``context'' in computing. He illustrates how discussions of ``context'' are better conceived as discussions of ``practices'' \cite{dourish}. Contextual information about a scientific observation or experiment is, indeed, a description of scientific practices: for example, the data collection process, the equipment used, the researchers involved in the observation, and the exact location of an observation. These types of contextual data are often only minimally described or left out entirely from the presentation of research results in scholarly publications, which are often all that researchers have when finding and using data collected by someone else. 

Researchers apply their own field expertise to evaluate data collected by others for relevancy and identify any potential problems \cite{zimmerman:2007}. Data digital libraries and repositories provide access to data, but as long as researchers must rely on often incomplete published reports of the data collection processes as data quality verification, data are of very little use for future research, as they cannot be interpreted. When these contextual data are not available, the reusability and reproducibility of scientific results is at stake. As recently noted by Stodden \cite{stodden}, data reusability and reproducibility of results are crucial to modern scientific communication, especially as scientific inquiry progressively becomes more dependent on scientific computation. Besides reproducibility, the availability of data and tools also ensures that further research on the same or other data sources can be conducted. 

Cultural norms of data sharing in different scientific communities are
intimately tied to the contexts of data production, namely the social
practices, relationships, and material conditions in which data
production takes place \cite{vertesi}. Because scientific practices
change significantly from domain to domain (and even from project to project), providing researchers with reliable tools and techniques to describe their data collection practices and the context of their research can be a challenge \cite{chin:2004,wallis_ijdc}. This challenge is even more marked in field based research that uses sensing technologies and Wireless Sensing Systems (WSS). With the addition of WSS, field research in areas such as environmental biology, seismology and urban sensing is becoming highly instrumented, computational, and collaborative \cite{cuff}. The utility of data collected \textit{in situ} with these systems, be it a natural reserve or an urban context, increases when coupled with contextual data that describe the setting in which the observations were taken: the equipment used to collect data, the researchers involved, and specific environmental conditions. Ensuring capture, description and preservation of these data is a fundamental task for scientific information management, as large volumes of sensor-collected data are linked to specific times and places and are thus irreplaceable. For these data, losing descriptive contextual information means losing the data altogether \cite{pepe:2007}. 

With these notions in mind, what tools can we provide scientists with to support collection of contextual data and documentation of the variegated artifacts that are produced in the scientific information lifecycle? In particular, can contextual data be collected in the field via handheld devices such as smart phones? These devices are now ubiquitous, contain many capabilities that might be useful for collecting contextual data, and can be equipped with  both custom and commercial software applications.  In this article, we discuss the use of microblogging, or short text annotations, as a mechanism to document the context and data collection practices of field-based research applications. We describe our design considerations of a handheld application and report on the results of interviews and a pilot study with researchers involved in \textit{in situ} ecological monitoring and related sensing research conducted in the field in natural and urban contexts.

\section{Scientific uses of microblogging}
Microblogging is a form of online communication by which users publish and broadcast content to the public or to a limited group of contacts. While \textit{blogging} allows the publication of lengthy, multi-media, user-generated content, \textit{microblogging} is limited to very brief, text-only messages. On the popular microblogging platform Twitter (\url{http://twitter.com}), microblog posts are commonly known as \textit{tweets}. Tweets are similar to SMS messages exchanged on cell phones, being 140 characters in length, at most. The brief format of tweets makes microblogging a timely and dynamic means of online communication. Microblogging is used to exchange and broadcast different types of information. 

Scholars from many disciplines---social, political, and computer scientists, for example---are increasingly becoming interested in the profusion of user-generated data available on microblogging sites. As more and more people share information about themselves via these social media, the emotional, political, cultural, and behavioral traits of entire communities begin to emerge. For example, researchers have recently drawn upon microblogging corpora to study news diffusion in geographic local networks \cite{dmb3}, and to model public mood levels against socio-economic indicators \cite{bollpepe}.

But scientists are not only \textit{analyzing} Twitter and similar microblogging platforms. They are also increasingly \textit{using} them. Scientists use Twitter to network with other scientists, to support the organization of scientific events, and to publicize their work. A recent survey conducted among computer scientists revealed Twitter to be among the top three services for dissemination of scientific information \cite{letierce}. In other related research, based on interviews and content analysis, it has been noted that Twitter has recently grown into a well-established vehicle of informal scientific citation \cite{priem}. Twitter has also been noted to be the current platform of choice to support the communication infrastructure of conferences: researchers use microblogging to produce aggregated real-time proceedings and to network with fellow attendees \cite{ebner:2009}. 

It is unclear, however, the degree to which the penetration of Twitter and other microblogging services in scientific practices  extends beyond the functions of social networking and information dissemination. Scientists use microblogging to publicize their work, exchange scientific information, cite interesting scientific work, and become acquainted with other scientists in their community. As we note in the next section, scientists also potentially can use microblogging and similar content publication services as an integral component of their scientific research activities, specifically to capture and document the context of highly-mobile field-based environmental and urban research. Are microblogging services used by field researchers to capture contextual information very early in the scientific data lifecycle, enabling them to document and adjust their research methods and data collection practices as they interact with each other, the experiment location, the data collection equipment, and other technical and social constraints?

\section{Why microblogging is convenient to document the context of
  field research data}
The work presented in this article is developed within the scope of the Center for Embedded Networked Sensing (CENS), a National Science Foundation Science and Technology Center involved in theoretical and applied work in the field of wireless communication networks and sensor technologies. CENS is a multi-institution research center which is comprised of five member universities in California: University of California, Los Angeles; University of Southern California; University of California, Riverside; California Institute of Technology; and University of California, Merced. CENS' goals are to develop, construct and apply sensing technologies to address questions in the natural environment (habitat ecology, marine microbiology, environmental contaminant transport, seismology) as well as in social and urban contexts. The type of research conducted at CENS is mostly conducted by teams of five to ten people in field settings and it spans a wide spectrum of disciplines and applications related to and supporting sensor network research. 

In sensor network research, a \textit{deployment} is a research activity in which sensors, sensor delivery platforms, or wireless communication systems are taken out into the field to study phenomena of scientific interest. CENS deployments have taken place at numerous locations in California (at various national reserves, lakes, streams, and mountains) and around the world (including Bangladesh, Central and South America). CENS researchers engaged in sensor deployments normally work in small teams in specific locales using newly developed instrumentation and create heterogeneous data sources. As such, they require flexibility to adjust research plans on the fly. The use of wireless sensing systems adds complexity to both the research activities and the kinds of information that researchers produce, because such systems are difficult to deploy in unpredictable field settings and often produce unreliable measurements.  

The goal of our research, in the Statistics and Data Practices Team at CENS, is to develop tools and applications that facilitate the collection and interpretation of data in scientific practices and field-based activities at CENS. In recent work, we have participated in the development of two digital data repositories: one for sensor-collected raw data, called SensorBase (\url{http://sensorbase.org/}) \cite{sensorbase}, and one for related contextual information, called CENS Deployment Center (\url{http://censdc.cens.ucla.edu/}) \cite{mm}. These two platforms are intended to function in parallel, as they represent different ``views'' of data relative to the same scientific lifecycle. SensorBase provides access to raw data and the Deployment Center provides enough contextual information to interpret them. More detailed information about the data hosted by these two repositories is shown in Table 1.

\begin{table*}
\centering
\caption{Side-by-side overview of the raw data and contextual data repositories}
\begin{tabular}{|l|p{7cm}|p{7cm}|} \hline
& SensorBase & CENS Deployment Center\\
\hline\hline
Hosted data& Raw and unprocessed data collected from sensors, semi-processed and analyzed data & Contextual data, descriptive information about field practices, equipment, times, and locations\\ 
&&\\[0.1pt]
Data types & Heterogeneous (ASCII tables, images, audio)  & Free text annotations, images\\
&&\\[0.1pt]
Metadata & Minimal, customizable; may vary by research field, application, and working group & May vary depending on the field, but time and location are usually recorded \\
&&\\[0.1pt]
Ingestion & Blogging & \textit{Micro-blogging?}\\
\hline\end{tabular}
\end{table*}

Raw data collected, produced, and managed by scientists and engineers performing WSS-based field research are extremely heterogeneous in nature. This is due both to the fact that sensor network research involves the interaction among researchers from a wide spectrum of domains with differing scientific data practices (e.g., biologists, seismologists, electrical engineers, computer scientists) and to the fact that different research applications necessarily produce different kinds of data. Data handled by CENS field researchers differ not only in media types (text, images, audio, video, software) but also in the conceptual role that these data have within a broader data lifecycle \cite{pepe:ore}. For this reason, Sensorbase has been designed and constructed as a central server with a flexible, and highly customizable metadata schema. Moreover, to encourage scientists to publish their data, SensorBase employs a light-weight data ingestion mechanism by which users upload and share data in a way functionally similar to the way by which bloggers post and share personal blog posts. 

The contextual data hosted on the Deployment Center are less heterogeneous in content and format than raw data are. They mostly describe the details of the setting in which field research is performed, such as information about field practices, equipment, times, and locations. The contextual information that CENS researchers create for their own use in their day-to-day work, however, exist in multiple media formats and are most frequently conveyed using free text annotations or images associated with a time and location. Researchers within CENS rarely create documentation that is not directly tied to their own use of their data, and correspondingly, they rarely share data with users from outside of their immediate projects \cite{matt}. With these notions in mind, we seek to determine whether micro-blogging is a viable ingestion mechanism for contextual data in field-based research. In other words, while blogging represents a convenient metaphor to figuratively represent the act of publishing and sharing raw data, is microblogging a valid metaphor to represent the act of publishing and sharing contextual data? In order to test the utility of microblogging as a mechanism for the ingestion and publication of contextual data we consider two crucial data challenges associated with most field-based research: interoperability and mobility. 

\begin{itemize}
\item \textit{Interoperability}. A conceptual and technical discrepancy exists among available metadata standards used to collect and represent contextual data in sensor network research. Locally-specific metadata representation formats for datasets and field notes often fail to inter-operate. In research areas new to computer-based instrumentation, few standards for description and annotation of resources such as sensor-collected ecological data are widely used. 

The challenge of interoperability could be addressed by the light\-weight nature of microblogging as a data posting service: microblog posts are brief and text-only. This means that microblogging services are fairly easy to implement in existing data collection platforms; also, collected microblog information is easily fed into existing data digital libraries. This also means that there is no need to create domain- or discipline-specific data models to accommodate different types of contextual information (e.g. contextual data collected in seismological projects and marine biology deployments). All contextual information embedded in microblogs can be collected and stored as text (and HTML links, in the form of text). Thus, microblogging could be an interoperable solution to document the context of field research practices. In turn, the availability of generic (i.e. domain-independent) contextual information in data digital libraries facilitates data sharing both within and across domains.

\item \textit{Mobility}. Ecological field deployments using wireless sensing systems necessarily involve in-field data creation, retrieval, and tracking. Understanding how high\-ly variable field-based methods impact the resulting data products is essential to understanding and interpreting data on both short and long terms. Many field locations where WSS deployments take place do not have Internet or local area network connectivity. Thus, approaches to collecting information about data collection activities for WSS deployments cannot be solely web-based. Mobile applications and devices may enable field researchers to perform basic data collection and management operations when working in remote locations where Internet connectivity is lacking. 

  The challenge of mobility could be addressed by the nature of microblog data: microblog posts natively carry metadata that is fundamental to contextual information, such as author and timestamp data---essential to reconstruct the link between the gathered contextual data and specific real-world situations. As discussed in the previous section, due to the unpredictable nature of mobile sensing research, losing, or failing to record, temporal information might make data collected in the field unusable. This is true, in most cases, not only for the collected sensor data but also for related contextual data. In a similar way, associating microblog posts with author information ensures data consistency and enables data reuse and exchange. Author and temporal metadata embedded in the gathered contextual information can be immediately published on the web, when researchers are in the field and a wireless Internet link exists, or can be stored locally on the device and later published on the web when Internet connectivity is lacking.
\end{itemize}

\section{Do field researchers use Twitter?}
Having argued that microblogging presents features that make it a
suitable mechanism to collect contextual data in field-based sensor
applications, we performed an exploratory survey among field
researchers to determine the current extent of adoption of Twitter and
similar microblogging platforms, be it for research or social purposes. We conducted interviews with eight scholars of different ranks (research staff and graduate students) involved in field research of various kinds (ecology, biology, environmental science, and urban sensing). Interviews were conducted in person or via email in a semi-structured format. The questions asked throughout the interview are the following:
\begin{enumerate}
\ii Have you ever used a micro-blogging service (e.g. Twitter)? If yes, what have you used micro-blogging services for? Do you use micro-blogging for research purposes?
\ii Do you know of other researchers in your discipline who use micro-blogging for research purposes? If so, what do they use micro-blogging for?
\ii Do you think that implementing a micro-blogging service on top of existing data collection tools/devices would assist you in your field work in any way? What specific information would micro-blogging facilitate to collect in the field?
\ii Do you think that micro-blogging could enable exchange of information (i.e., research-based conversations) with fellow researchers during field activities? If so, in what ways would that be useful for you?
\end{enumerate}

When asked about their micro-blogging practices, nearly all interviewed researchers said that they either do not use Twitter (or any similar micro-blogging platform), or if they do, their use is limited solely to social interactions. Only one researcher actively employed Twitter both for social and research purposes. This use is specifically limited within the scope of \textit{Budburst}, a citizen science mobile sensing project aimed at collecting data on phenology and climate change (\url{http://budburst.org/}). Project participants contribute to this project by performing multiple observations of the same plants, documenting  ``phenophases'' such as their first leaf, the first flower, and the first fruit of trees, shrubs, flowers, and grasses. There are multiple ways to make and record budburst observations, including sms text, a smartphone application, and Twitter. On Twitter, users mark tweets about observations with a given hashtag (\#budburst). The application scrapes the tweet corpus daily and a stream of observations becomes available to users and researchers on the project. The same researcher also indicated another research use of Twitter, which is not directly field-based, but that is related to the aforementioned project. In what they jokingly referred to as ``opportunistic'' participatory sensing, they continuously search the public Twitterverse for phenology-related keywords, such as ``spring'', ``bloom'', ``blossoming'', etc. They log tweets containing these key terms and their geotag, if available. Analysis of this corpus allows them to correlate user-contributed Twitter events with life cycle events in plants.

All the researchers that do not use Twitter also indicated that they do not know of any colleagues who do use it for research purposes. Some researchers were aware, however, of some field-based research applications that use aspects of blogging or micro-blogging as their data ingest mechanism. One researcher mentioned the existence of a Facebook application for beach contamination source tracking. In the realm of urban sensing, a researcher mentioned a project of a colleague who uses the distribution of tweets across cities to look at urban landscapes and urban density. Nearly all interviewed researchers indicated that the current most common and beneficial use of Twitter for research purposes has to do with scientific communication, dissemination of scientific results, and gaining visibility in one's scientific community:

\begin{quote}
\begin{small}
[Student (Biology)] I know some researchers who seem to use Twitter
for informing their ``followers'' of their goings-on, which often
include what research/field campaigns they're working on.. I've also
seen a fair bit of micro-blogging of ``popular'' research, i.e. links
to fun, kinda cool research...
\end{small}
\end{quote}

\begin{quote}
\begin{small}
[Student (Ecology)] What I could picture for me personally as a graduate student is maybe augmenting my personal research website, so I could see using Twitter to maybe update people on the status of my research, and using that application to do that. I guess it would be a way to gain a wider audience and feedback for my research, in real time. I see that as a potential application.
\end{small}
\end{quote}   
 
Researchers indicated a number of ways by which microblogging services could assist their field work practices. One researcher suggested the use of Twitter feeds for automatic notification of instrument failures. Device failure is a problem that spans across all mobile sensing applications. The use of a Twitter feed, in addition to email notifications, would allow researchers to instantaneously subscribe to multiple devices, re-broadcast (retweet) given failures to a broader/different research community, and publish notes and annotations concomitant with the failure messages, for public viewing. One respondent mentioned the potential of microblogging platforms to foster public awareness and interest in scientific research. For example, it was suggested that Twitter feeds could be useful to publish water quality data about particular beaches. This in turn, has the potential to serve the research community---other researchers might be interested in harvesting, storing and reusing those data---as well as the public---citizen scientists and beachgoers may find interest in the data and in the processes by which those data were collected.

Some researchers pointed to an important shortcoming of Twitter: its inability to function in field locations in which Internet connectivity is not available. As discussed above, the mobile and unpredictable nature of most scientific field research implies that Internet connectivity may be lacking for extended periods of time. If Twitter had to be used as a note-taking tool for the collection of contextual data, how would it behave in the absence of Internet connection? The doubts expressed by interviewed researchers point to the importance to employ microblogging services with offline capabilities when performing field experiments.

With regard to the last question of the interview, many researchers were unsure whether microblogging is a viable platform for exchange of research-based information during field activities, for two main reasons: the potential lack of connectivity in field settings and the public nature of Twitter. The connectivity issue was raised as a major cause for the inability to communicate while in the field:

\begin{quote}
\begin{small}
[Research staff (Ecology)] Again, there is the connectivity issue. It's not clear how this would be better than walkie-talkie, or just saying, ``lets all meet here at \_\_\_''. I have been using my twitter account to post about the progress of projects, but these are just general posts, not field work specific. 
\end{small}
\end{quote}   

Other researchers were worried about the public nature of Twitter, and thus the issue of publishing raw, non-certified field data into the public: 
\begin{quote}
\begin{small}
[Student (Biology)] I suppose micro-blogging could foster collaboration between researchers while in the field, but with the public nature of things like Twitter I don't know that it would be appropriate for sharing data or results of field work prior to any QA/QC processes.
\end{small}
\end{quote}   
Although Twitter does allow accounts to be private, i.e., to be accessible only by a selected group of contacts, researchers expressed the need to make openly available certain sections of their Twitter feeds, while restricting others.

In sum, although all researchers in our study possess some familiarity with Twitter, very few actually employ microblogging services actively. Some of them use Twitter for promotion of their scientific work and to network with other scientists. Although most interviewed researchers consider feasible and convenient the use of Twitter to support field-based research, only one of them has personally experimented in this regard, collecting field data about phenophases and climate change. Two major shortcomings of Twitter were put forward: its inability to work in offline mode and its limited range of options on access privileges. Combining these findings with the fact that CENS researchers largely create documentation that is directly tied to their own use of their data \cite{matt} suggests that microblogging has little added value for field researchers in-and-of-itself as a research tool that exists in isolation from other research tools and the data themselves. In the next section, we present a mobile application that tries to address this problem by bringing the microblogging service closer to the data.

\section{System design and development}
In light of interview results and ongoing participant observation with field researchers, we outline the requirements and sketch the design of a microblogging-like system that would allow collection of data and field contextual metadata together. As we discussed, researchers engaged in sensor-based field work perform research in a very mobile manner, they work in small teams using ad-hoc methods, they produce heterogeneous data sour\-ces, and require enough flexibility to adjust research plans and data ingestion mechanisms. In designing, deploying and testing a data collection tool for this community, we focus on enabling flexible and adaptive research. 

Our design and code base derived from another CENS project, the EcoPDA
project\footnote{\url{http://www.teamnetwork.org/en/content/ecopda}}. The
EcoPDA application provided a handheld system with which to conduct
biodiversity surveys. Our application sought to develop a more
general-purpose system which could be used in any type of field
research. We also drew design ideas from other similar projects.  The
ButterflyNet project\footnote{\url{http://hci.stanford.edu/research/butterflynet/butterflynet/}} was another project to develop a handheld application for collecting field research data, but required specialized paper and pens to use. The Cybertracker project\footnote{\url{http://www.cybertracker.org/}} provides another handheld tool for performing field research, but again is designed to be used to perform biodiversity surveys, and as such is not as general purpose as we would like our application to be.

Our development follows a \textit{usability engineering lifecycle}, which organizes the design process around two phases: (i) requirements analysis and (ii) design, testing, and development \cite{mayhew}. In the first phase, requirements analysis, we developed a user profile based on our discussions with field researchers. We performed contextual analysis of the users' main work tasks, set usability goals, analyzed the platform capabilities and constraints, and created general design principles. Field research in the ecological and environmental sciences takes on many forms depending on the research questions being asked. In recent sensor network deployments, we have observed that scientists and engineers involved in field work collect in their notebooks many diverse kinds of contextual data such as information relative to the equipment used, participants of the project, annotations of various types, and a number of tasks performed during, before and after the field work \cite{mm}. Some of this information is very domain-specific. For example, researchers involved in the study of environmental sensing and image acquisition collect information that indicates the detailed models and characteristics of the imaging sensors installed and used. Researchers working in urban sensing field projects collect and store entirely different contextual data, such as social and demographic information relative to the urban environment being studied. Although disparate kinds of contextual data are being collected, there are certain categories of field activities that occur with regularity across all domains and practices of field-based research. This allows us to focus our design and development efforts on specific categories of field activities, yet leaving the system sufficiently open to enable user appropriation, i.e., researchers may progressively use the provided microblogging functionality for uses that are outside the initially intended use of field-based note taking.

Our design requirements and principles stem from our interviews as well as from ongoing and past participant observation of field research activities at CENS and can be summarized as follows:
\begin{itemize}
\item \textit{Provide full metadata customization and flexibility}: In the field, highly varied kinds of data are produced. We would like to give researchers full control over the metadata required in the documenting the context of these data. Some researchers might want to create pre-existing metadata schemes. Others might want to produce solely unstructured free-text annotations.
\item \textit{Specify data collection protocols}: When researchers cho\-ose to specify their own collection procedures, they have to be able to fully customize the data collection interface to those procedures.
\item \textit{Collect repeatable data}: Researchers often repeat data collection procedures, both to replicate previous results and to augment prior experiments. The system must allow them to re-use data collection protocols that they have already created and used.
\item \textit{Perform their field activity as fast with the application as without}: Time spent in the field is precious. If our system noticeably slows researchers down, its usefulness is greatly diminished. 
\item \textit{Easily integrate the field data collected with the tool into their existing data collections}: Researchers in most sciences are already facing challenges in organizing and using digital data resources, thus our system must integrate with existing data and data tools.
\item \textit{Provide both online and offline ingestion and access}: Researchers must be able to both access and populate the system (i.e., produce microblogging annotations) from any location (their office, or the field) whether they are online or offline. 
\item \textit{Allow both public and private publication of content}: Some content, such as preliminary observations, sensitive material, and equipment information, is only intended for a close circle of researchers, directly involved in a field project. Other content can be safely made open and publicly available. The system must allow users to publish content in both public and private mode, and to quickly switch between these two modes, if needed. 
\item \textit{Produce data annotations that ``live with'' the data}: Pa\-ckaging data and its contextual information increases the usefulness, and consequently the value, of the data itself. Thus, we would like to develop a microblogging service that not only produces annotations, but it also gives users the possibility to associate those annotations with specific datasets subsets of data, and data points. 
\end{itemize}

The second design phase---design/testing/development---involves a number of iterative steps, including: conceptual modeling, conceptual mockups, setting screen design standards, prototyping, evaluation of the standards and prototyping, interface design, and evaluation of the interface design by gathering user feedback. 

\begin{figure}[h!]
\centering
\includegraphics[width=0.43\textwidth]{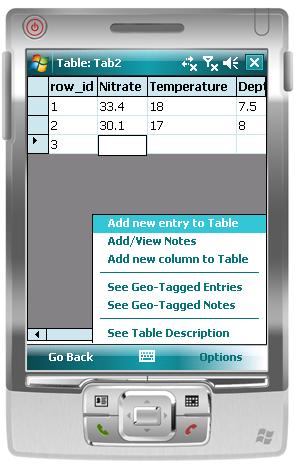}
\caption{\label{fig:1} A screen capture of the mobile phone application for the collection of tabular data.}
\end{figure}

The focus of our development work thus far has centered around the development of a Twitter-like annotation platform that would integrate with existing data collection tools and instruments employed in the field. One such tool is an application which runs on a handheld device and allows field researchers to collect tabular data. The application, developed on Windows Mobile 6.1 operating system and currently tested on Samsung Omnia smart phones, features a simple web interface by which users can ``author'' data collection tables prior to going out in the field, i.e., they have the option to structure a basic metadata schema a priori. Once in the field, these data collection tables are customizable, in that they can be changed and re-used, facilitating data repeatability. Figure 1 shows a screen capture of a data collection table on the mobile phone. The menu on the bottom right shows the options available to the user, which include standard options such as adding an entry (data point), and adding columns to the table.

\begin{figure}[h!]
\centering
\includegraphics[width=0.43\textwidth]{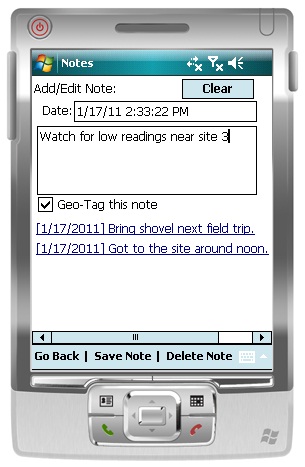}
\caption{\label{fig:1} A screen capture of the note taking interface for the collection of contextual data.}
\end{figure}

Following the list of requirements and principles outlined above, we developed a note taking service that operates within this data collection tool. The service gives users two additional options: to add/view notes and display the geo-tagged items, as shown in the menu on the bottom right of Figure 1. When using these functionalities, a user is presented with a note taking interface, shown in Figure 2. Importantly, since the annotation service can be invoked directly from within a specific data table, annotations can be made data-specific, i.e. directly associated with a given dataset or data table. The initial prototype only allows annotation at the level of data tables, but in subsequent development we plan to extend the annotation capability to specific data rows, columns, and points. For example, in the case of Figure 1, if the menu is invoked under \texttt{row\_id} 3 and column \texttt{Nitrate}, the relative annotation will be stored specifically within the scope of that data table. This allows annotations to ``live with'' the data, i.e., full integration of annotation functionalities with existing data and services, by being stored in the same XML output file. The note taking interface we developed, shown in Figure 2, is not very different from typical microblogging content publishing services. It is composed of the following elements:
\begin{itemize}
\ii \textit{Date- and time-stamp}: The service automatically populates date and time fields, which are stored alongside the text annotation. This can be altered if needed (e.g., if the researcher wants to make a note about a past event or set a reminder about a future event).
\ii \textit{Text box}: This box contains the content of the annotation. Unlike Twitter, the maximum allowed count is not limited to 140 characters.
\ii \textit{Geo-tagging functionality}: Users can geo-tag data po\-ints and notes by checking the ``Geo-tag this'' box. The geo-tag function pulls the latitude and longitude of the user's current location from the phone's built in GPS receiver and links the coordinates to the data point or note that was tagged. The user can then filter the display to view the data points and notes that have been geo-tagged\footnote{It has been noted that in many situations the GPS coordinates are of little meaning or use in and of themselves \cite{sorvari}. To address this problem, we are exploring ways to visually present the geo-tagged information, for example through a map interface.}. Also, devices that have GPS capability may not be usable in field locations because of physical obstructions. In these cases, users can self-describe the location using the free-text note field. Although geographically less precise than GPS, these descriptions might provide some context when GPS data are entirely missing.  
\ii \textit{Annotation feed}: the lower portion of the interface provides a list of recent annotations in reverse chronological order, similar to the Twitter news feed.
\end{itemize}

Not shown in the images are two other components of the note taking application which reflect our design considerations. First, the service does not depend on an Internet connection to function. It stores all annotations locally until they are pushed, at given time intervals, to a desired online database, or microblogging service. When no Internet is available, all stored annotations are stored and queued in the system and then pushed online as soon as a connection becomes available. Second, we have only partly addressed the problem relative to access privileges. As discussed above, users should be able to decide whether to publish certain annotations publicly or with a restricted group of contacts. For the time being, we have addressed this problem in the following manner. All annotations produced are pushed by default to an online private microblogging database that is only accessible by CENS field researchers. However, our planned approach is that upon submitting an annotation, users will be asked whether they also want to push the annotation to a public microblogging service (e.g., a Twitter account), or to Sensorbase and/or the CENS Deployment Center. Using these mechanisms, field researchers can make annotations without worrying about the availability of a wireless Internet connection. They also can be certain that all their notes are eventually pushed to a private online database, and in addition will be able to choose which ones are pushed to an online microblogging platform for public viewing. And finally, because we noted that a number of interviewed researchers use Microsoft Excel formats for their day-to-day data collection, manipulation, and analysis, we allow users to export their data and notes in an Excel file, again enabling the data to ``live with" the contextual notes. The Excel output allows users to integrate their data collected via our application with their existing data collected via other means.

In the next section, we present a use case scenario of the application together with the results of preliminary user pilot study. 

\section{Pilot study}
Our first round of pilot users consisted of three scientists with extensive experience in performing environmental and ecological research in field settings. The pilot tests took place indoors in a laboratory setting, and focused on the usability and utility of both the data collection and note taking applications described in the previous section. The pilot test protocol consisted of five tasks aimed at testing the main functionalities of the system, and how easy the application was to learn and use. First, we had the users test the web interface that allows them to create data collection tables, and second, we had them test the application using these tables. This involved opening the tables, taking data, customizing the tables by creating new columns, taking notes, and finally exporting the data. In this stage, we did not help users in any way, to mimic a real-world environment as much as possible. We asked users to think aloud as they completed the tasks in order to gain a better understanding of their thought processes.  

The feedback from our pilot users highlighted some positive and negative aspects of our prototype system. Users liked the ability to create their own data tables, and the ability to customize the tables during the data collection process, by adding new columns. They also liked the mobile aspect of the application, in that the small size of a mobile phone will not be an encumbrance as they perform their research activities in varied field settings. The small size of the application had both positive and negative aspects, however. Both users reported that the small screen size of the phones might prove to be hard to see in some situations, specifically mentioning button size and text size as potential issues. We also identified individual pages and functions that were either hard to use (such as the data import function) or did not work as we had anticipated. This enabled us to see where the application needed to be streamlined or changed in order to provide smoother user experience. 

The other main motivation of the pilot test was to get feedback on the note\-taking capabilities of the mobile data collection application. All three pilot users are not active Twitter users. One user said that they had previously utilized smartphones to take and post pictures during field research, but none of them currently uses Twitter or any other microblogging service for field research in any major way. Pilot users were invited to utilize the note taking device. They noted that this application would be most useful for taking notes on the spot while in the field. One tester stated that this meant that they would not have to carry a notebook around, and it would allow them to attach the location information to their notes. All users agreed that the collection of notes as short text messages and their preservation and distribution via a database or website would be very useful. Two users also suggested that the note taking application is most useful in situations that involve collecting information about events. More specifically, sending notes out as microblogs is useful in providing notifications to other members of the research team or to the general public that a certain event has taken place, or that important thresholds in data collection have been achieved. All users also mentioned that attaching GPS or other location information to the annotations certainly increased their information utility.

\section{Conclusion}
Throughout this article, we have stressed the importance of contextual information and data in understanding and using information resources in any setting, from single-investi\-gator projects to distributed and collaborative projects. In particular, we focus on the challenges related to collecting contextual information while performing research in the field, especially when using wireless sensing systems in unpredictable and ever-changing field settings. We have suggested that using microblogging as a text annotation tool might allow the ingest and publication of contextual data from the field because microblogging offers a convenient platform to mimic some of the annotation functions of a notebook. 

Results of our interviews with field researchers suggests that microblogging is not widely used as a field research tool currently, but that potential exists for microblogging to be widely adopted as a de facto field annotation tool, at least for two reasons. First, microblogging is already a widely popular and used form of scientific communication and networking on the web. Researchers are familiar with microblogging tools, and the ubiquity of these tools across social settings and technical platforms ensures that they will be widely available for adoption and appropriation. Second, certain aspects of microblogging accommodate the highly mobile nature of field research, specifically, its broad range of uses and applications, and the heterogeneous nature of data produced. Yet, we identified certain shortcomings of Twitter and similar microblogging platforms: their inability to work in offline mode, their limited range of access privilege options, and their current isolation from other data collection and annotation tools.

Based on our interview results and participant observation with a community of ecological, environmental, and urban sensing researchers involved in field work, we presented the design and development of a microblogging-inspired application for the collection of observational and contextual data from field research activities. We discussed how our application accommodates field research by addressing the aforementioned challenges and by integrating directly with the data and existing data collection tools. Finally, we ran a preliminary pilot test to evaluate the utility and usability of the tabular data and note taking applications. While the pilot test was very limited in scope and scale, it was useful to gather expert feedback from field scientists on key functionalities of our application, and shape ideas for the next steps. A fuller evaluation of the application will be conducted at a later date, in real-world settings. 

This study suggests that the use of microblogging services in existing
handheld and other in-field data collection devices can prove an
interoperable and mobile tool to produce and publish contextual data
in the unpredictable, variable settings of highly-mobile field
research, but, crucially, only if the microblogging tools are
integrated within the larger set of field researchers' technologies
and practices. Collecting contextual data about scientific artifacts
is a continual challenge in digital library and repository creation
and adoption. We envision that researchers using microblogging tools
as part of their field research activities will engage in patterns of
interactivity (e.g. sharing annotations, subscribing to each other
annotation feeds, etc.) that are ultimately beneficial for the pursuit
of collaborative field research, in that these annotations and
collaborative feeds can serve as a source for contextual information
about data, publications, field notes and images, technical reports,
and any number of other digital resource genres that are and will be
collected and made available in digital libraries. We hope that
widespread use of microblogging and related annotation tools in field
research will enrich scientific data repositories with important contextual information, thus enabling reuse and interpretation of data collected in field activities.



\bibliographystyle{IEEEtran}
\bibliography{alberto}

%



\end{document}